\documentclass[10pt,english,reprint,showpacs,showkeys]{revtex4}
\usepackage[T1]{fontenc}
\usepackage[latin9]{inputenc}
\usepackage{graphicx}
\usepackage{amssymb}
\usepackage{esint}

\makeatletter

\providecommand{\tabularnewline}{\\}

\@ifundefined{textcolor}{}
{%
 \definecolor{BLACK}{gray}{0}
 \definecolor{WHITE}{gray}{1}
 \definecolor{RED}{rgb}{1,0,0}
 \definecolor{GREEN}{rgb}{0,1,0}
 \definecolor{BLUE}{rgb}{0,0,1}
 \definecolor{CYAN}{cmyk}{1,0,0,0}
 \definecolor{MAGENTA}{cmyk}{0,1,0,0}
 \definecolor{YELLOW}{cmyk}{0,0,1,0}
 }

\@ifundefined{showcaptionsetup}{}{%
 \PassOptionsToPackage{caption=false}{subfig}}
\usepackage{subfig}
\makeatother

\usepackage{babel}

\begin{document}

\title{Theoretical model to deduce a PDF with a power law tail using Extreme
Physical Information}

\author{Ricardo Bonilla}

\email{rbonilla@uniandes.edu.co}


\author{Roberto Zarama}

\email{rzarama@uniandes.edu.co}

\affiliation{Universidad de los Andes, Bogotá, Colombia, Ceiba-Complejidad}

\author{Juan Alejandro Valdivia}

\email{alejo@macul.ciencias.uchile.cl}

\affiliation{Departamento de Física, Facultad de Ciencias, Universidad de Chile,
Santiago, Chile}
\begin{abstract}
The theory of Extreme Physical Information (EPI) is used to deduce
a probability density function (PDF) of a system that exhibits a power
law tail. The computed PDF is useful to study and fit several observed
distributions in complex systems. With this new approach it is possible
to describe extreme and rare events in the tail, and also the frequent
events in the distribution head. Using EPI, an information functional
is constructed, and minimized using Euler-Lagrange equations. As a
solution, a second order differential equation is derived. By solving
this equation a family of functions is calculated. Using these functions
it is possible to describe the system in terms of eigenstates. A dissipative
term is introduced into the model, as a relevant term for the study
of open systems. One of the main results is a mathematical relation
between the scaling parameter of the power law observed in the tail
and the shape of the head. 
\end{abstract}

\keywords{Fisher Information, power law, fitting long tail, head distribution
analysis, Extreme Physical Information}

\pacs{89.70.Cf, 89.75.Da, 02.60.Ed, 04.20.Fy}

\maketitle

\section{Introduction}

In the natural and social sciences there are many variables that distribute
as power laws \cite{Newman2005,Andriani2009}. Several models are
described in the literature to explain and fit these power laws. There
are models from first principles \cite{Mandelbrot1953,Bak1987,Abe2000,Yang2004,Pennini2004},
microscopic interactions \cite{Barabasi2002,Mitzenmacher2002,Bohorquez2009},
using stochastic models \cite{Reed2003,Lammoglia2008} and even ad
hoc mathematical formulas \cite{Sarabia2009}. The problem identified
in many of these models is the lack of analysis of frequent events
in the distribution head and the possible relations between head events
and extreme events in the tail. The aim of this paper is to construct
a model that simultaneously describes the power law in the tail and
the shape of the head. 

Reed \cite{Reed2003} has proposed a model that describes and fits
the head and tail of a social variable. Reed\textquoteright{}s model
is stochastic with independent parameters for head and tail. This
article describes a model based on basic information principles, from
which a relation between the head and the tail parameters is derived.

The model builds on the basic minimum information principle, using
Extreme Physical Information (EPI) theory \cite{Frieden2004}. EPI
theory is used in a simple and particular case to derive a second
order system differential equation. This equation of motion reveals
the phenomenon dynamics and its solution is a probability density
function (PDF) with a power law tail. As an extension to the model
a dissipative term is proposed in order to take into account that
real systems are usually open, exchanging matter, energy, or information
with their surroundings.

Due to the fact that the proposed model is theoretic, it can be used
to infer the possible PDF governing a system when its microscopic
characteristics are known. The model can also be used to fit an observed
distribution numerically. As will be shown later when fitting the
tail of an observed distribution, the shape of the head is determined
via an equation that relates the scaling parameter in the tail to
the shape of the head.

This paper is organized as follows: first, the Fisher information
measure is presented; by applying EPI theory, then an information
functional is constructed, solved, and the PDF of the system computed;
next, a dissipative term is introduced into the information model;
finally, the model is used to fit the head and the tail of 4 different
power law tails that have been observed and already studied in the
literature.

\section{The information model}

The model is based on EPI theory. This theory was developed by Roy
Frieden, building on the basic information measure of Fisher information
\cite{Frieden1990,Frieden1996}. In EPI theory, this measure is used
to construct an information functional that is minimized to get an
equation describing the motion of the system.

In the literature EPI theory has been used to deduce several physics
equations; for example, it was used to derive the Schrödinger equation
in quantum mechanics and Einstein\textquoteright{}s field equations
of general relativity \cite{Frieden1995}. It was also used to derive
classical statistical physics in thermodynamics \cite{Frieden1999}.
The EPI theory has also been applied to biological, economic and other
complex systems, with interesting results \cite{Binder2000,Gatenby2002,Hawkins2004,Frieden2005,Gatenby2007,Hawkins2010}.
In this paper, EPI theory is used to derive an analytic PDF of a system
with a power law tail. The deduced PDF is a piecewise function with
two parameters, one for the distribution head and the other for the
tail. The major contribution of this model is to show that the two
parameters are dependent.

\subsection{Fisher information}

The information measure used in EPI theory is Fisher information.
This measure was derived in 1922 by R.A. Fisher from estimation parameter
theory \cite{Fisher1922,Fisher1925}. In estimation parameter theory,
an estimator $\hat{\Theta}(x)$ of an unknown quantity $\Theta$ is
unbiased if

\begin{equation}
\left\langle \hat{\Theta}(x)-\Theta\right\rangle \equiv\int dx\,[\hat{\Theta}(x)-\Theta]f(x|\Theta)=0,\label{eq:UnbiasedEstimator}\end{equation}
where, $f(x|\Theta)$ is the PDF of the observed variable $x$ given
a parameter $\Theta$. This equation yields to the Cramer-Rao inequality\begin{equation}
e^{2}\geq\frac{1}{I},\label{eq:Cramer-Rao}\end{equation}

where, $e^{2}$ is the mean-square error and $I$ the Fisher information.
This relation means the error of a measurement is bounded by the Fisher
information. Developing Eq. (\ref{eq:UnbiasedEstimator}), Fisher
information can be written as\begin{equation}
I=\int dx\,\frac{1}{f(x|\Theta)}\,\left(\frac{\partial f(x|\Theta)}{\partial\Theta}\right)^{2}.\label{eq:FisherInfo}\end{equation}

When the variable $x$ obeys the shift relation $x\equiv\Theta+y$,
it is possible to simplify Eq. (\ref{eq:FisherInfo}). If $g(y|\Theta)$
is the PDF of $y$ given $\Theta$ then

\begin{eqnarray}
f(x|\Theta) & = & f(y+\Theta|\Theta)\nonumber \\
 & = & g(y|\Theta).\label{eq:Shitf-Relation}\end{eqnarray}

Finally, if random variable $y$ was independent of the size of $\Theta$,
Eq. (\ref{eq:Shitf-Relation}) becomes\begin{equation}
f(x|\Theta)=g(y).\label{eq:Shift-Relation02}\end{equation}

Using Eq. (\ref{eq:Shift-Relation02}) in Eq. (\ref{eq:FisherInfo})
the Fisher information simplifies to \begin{equation}
I=\int dy\,\frac{g'(y)^{2}}{g(y)},\label{eq:FisherInfoSIMPLE}\end{equation}

where $I$ represents the information retrieved through the measurement
of variable $y$. The goal of EPI will be the computation of $g(y)$
by means of an information balance.

\subsection{Application of EPI }

EPI theory involves the construction of an information functional
to be minimized. The functional must have at least two terms. The
first term is always the Fisher information Eq. (\ref{eq:FisherInfoSIMPLE})
and the second term is the bounded information. The Fisher information
is the amount of information retrieved when parameter $\Theta$ is
estimated measuring $y$; the bounded information is a term that can
be constructed via a unitary transformation of the Fisher information.
Here, the unitary transformation will be the conservation of the probability
between the PDF $g(y)$ and a second PDF $h(z)$, given that $y$
is a function of $z$.

The model developed in this paper presupposes the existence of two
observable variables, $y$ and $z$, that are related by a the linear
relation

\begin{equation}
z=t(y)\, y,\label{eq:relacion-lineal}\end{equation}
where, $t(y)$ is the piecewise function

\begin{equation}
t(y)=\left\{ \begin{array}{ccl}
t_{1}, &  & 1\leq y\leq y_{0}\\
t_{2}, &  & y>y_{0}\end{array}\right.,\label{eq:piecewiseF}\end{equation}

and where, $t_{1}\neq0$ and $t_{2}\neq0$ are constants.

If Eqs. (\ref{eq:relacion-lineal}) and (\ref{eq:piecewiseF}) were
known, then measuring $y$ or $z$ would produce a minimum information
discrepancy. It turns out that the information functional $\mathcal{F}$
to minimize is the Fisher information $I^{(y)}$, the information
retrieved when $y$ is measured, minus the Fisher information $I^{(z)}$,
when $z$ is measured:

\begin{eqnarray}
\mathcal{F} & = & I^{(y)}-\kappa I^{(z)}.\label{eq:Functional}\end{eqnarray}

The constant $\kappa$ is a variable that measures the information
discrepancy between measuring $y$ or $z$ in the system. Using Eq.
(\ref{eq:FisherInfoSIMPLE}) to write each term of Eq. (\ref{eq:Functional}),
the information functional becomes

\begin{equation}
\mathcal{F}=\int dy\,\frac{g'(y)^{2}}{g(y)}-\kappa\int dz\,\frac{h'(z)^{2}}{h(z)}.\label{eq:Functional-FI}\end{equation}

To minimize the functional $\mathcal{F}$, the second term on the
right of Eq. (\ref{eq:Functional-FI}) is written in terms of $g'(y)$,
$g(y)$ and $y$:\begin{equation}
I^{(z)}=\int dy\,\frac{1}{t(y)^{2}}\frac{g(y)}{y^{2}},\label{eq:Iz}\end{equation}
as shown in appendix \ref{sec:ABoundedInfo}.

Using Eq. (\ref{eq:Iz}) the information functional becomes

\begin{eqnarray}
\mathcal{F} & = & \int dy\,\mathcal{L}[g'(y),g(y),y],\end{eqnarray}

where

\begin{equation}
\mathcal{L}[g'(y),g(y),y]=\frac{g'(y)^{2}}{g(y)}-\frac{\kappa}{t(y)^{2}}\frac{g(y)}{y^{2}},\label{eq:Lagrangeano}\end{equation}

is the Lagrangian of the system. The solution for $g(y)$ is deduced
applying Euler-Lagrange equations. To simplify further calculation,
a change of variable is used,

\begin{equation}
g(y)=q(y)^{2}.\end{equation}

So, the Lagrangian becomes

\begin{equation}
\mathcal{L}[q'(y),q(y),y]=4q'(y)^{2}-\frac{\kappa}{t(y)^{2}}\frac{q(y)^{2}}{y^{2}},\end{equation}

with solution

\begin{equation}
4q''(y)+\frac{\kappa}{t(y)^{2}}\frac{q(y)}{y^{2}}=0.\label{eq:EqMotion}\end{equation}

Eq. (\ref{eq:EqMotion}) represents the equation of motion of the
system, and represents, therefore, the system behavior.

The general solution of Eq. (\ref{eq:EqMotion}) is piecewise because
function $t(y)$ is also piecewise, so: 

\begin{equation}
q(y)=\left\{ \begin{array}{ccl}
q_{1}(y), &  & 1\leq y\leq y_{0}\\
q_{2}(y), &  & y>y_{0}\end{array}\right..\label{eq:Q1Q2}\end{equation}

The general form for solutions $q_{1}(y)$ and $q_{2}(y)$ is\begin{equation}
q_{i}(y)=c_{i,1}y^{\frac{1}{2}+k_{i}}+c_{i,2}y^{\frac{1}{2}-k_{i}},\label{eq:EqMotionSolution1}\end{equation}

where $c_{i,1}$ and $c_{i,2}$ are integration constants and coefficient
$k_{i}$ obeys

\begin{equation}
k_{i}=\frac{1}{2}\sqrt{1-\frac{\kappa}{t_{i}^{2}}}.\label{eq:SimplificacionK}\end{equation}

The value of $t_{i}$ depends on the range of $y$ through Eq. (\ref{eq:piecewiseF}).
The equivalence with the solution of a free quantum particle in quantum
mechanics is straightforward, as shown in appendix \ref{sec:BDissipativeTerm}.

In section \ref{sec:DissipativeTerm}, Eq. (\ref{eq:EqMotion}) will
be complemented with a dissipative term to describe information flow
between system and environment.

\section{A particular solution\label{sub:AParticularSolution}}

To apply the model it is necessary to establish several boundary conditions.
These conditions are usually taken from the real system under study.
As an example, suppose a piecewise function $t(y)$ with one step,
as in Eq. (\ref{eq:piecewiseF}). Also, suppose that $q(y=1)=0$,
$q(y\rightarrow\infty)\rightarrow0$ and $k_{1}\in\mathbb{C}$. The
restriction over $k_{1}$ makes it possible to write $q(y)$ in terms
of Sines and Cosines, that is, in terms of an oscillatory wave. Obeying
the piecewise function $t(y)$, the solution for $q(y)$ is,

\begin{equation}
q(y)=\left\{ \begin{array}{lcl}
c_{1\,}y^{\frac{1}{2}}\sin(k_{1}\log(y)), &  & 1\leq y\leq y_{0}\\
c_{2}\, y^{\frac{1}{2}-k_{2}}, &  & y>y_{0}\end{array}\right.,\label{eq:QSolution}\end{equation}
where the power law in the tail may be identified. The PDF of the
tail can be written in terms of the typical Pareto distribution with
scaling parameter $\alpha_{tail}$ as\begin{equation}
q_{2}(y)^{2}=\left|c_{2}\right|^{2}y^{1-2k_{2}}=\left|c_{2}\right|^{2}y^{-(\alpha_{tail}+1)},\end{equation}

with \begin{equation}
k_{2}=\frac{\alpha_{tail}+2}{2}.\label{eq:k2-tail}\end{equation}

To compute coefficients $c_{1}$ and $c_{2}$ in Eq. (\ref{eq:QSolution}),
the continuity of $q(y)$ and $q'(y)$ at $y=y_{0}$ is used. These
conditions make it possible to calculate the two coefficients in terms
of $y_{0}$, $k_{1}$ and $k_{2}$.

Taking into account that the PDF must be normalized,

\begin{eqnarray}
\int q(y)^{2} & dy= & 1,\label{eq:Normalization}\end{eqnarray}

an additional equation is deduced\begin{equation}
k_{1}=-k_{2}\tan\left(k_{1}\log\left(y_{0}\right)\right).\label{eq:QuantizationFunction}\end{equation}

Eq. (\ref{eq:QuantizationFunction}) establishes a relation between
the scaling parameter of the power law tail and the shape of the head.
Using the normalization condition a relation between $k_{1}$ and
$k_{2}$ may be derived as a function of $y_{0}$. The relation will
be particular to each system, often a transcendental equation with
multiple solutions for $k_{1}$, given the value of $k_{2}$ and $y_{0}$.
In such cases, when the equation gives several solutions for $k_{1}$,
solutions can be seen as eigenstates of the system. In few words,
given a scaling parameter of the tail and the minimum value $y_{0}$,
it is possible to establish several eigenvalues for the shape of the
head: the tail of the PDF fixes the head. Up to now, this relation
has not been identified or studied analytically.

\begin{figure}
\subfloat[]{\includegraphics[width=0.4\columnwidth]{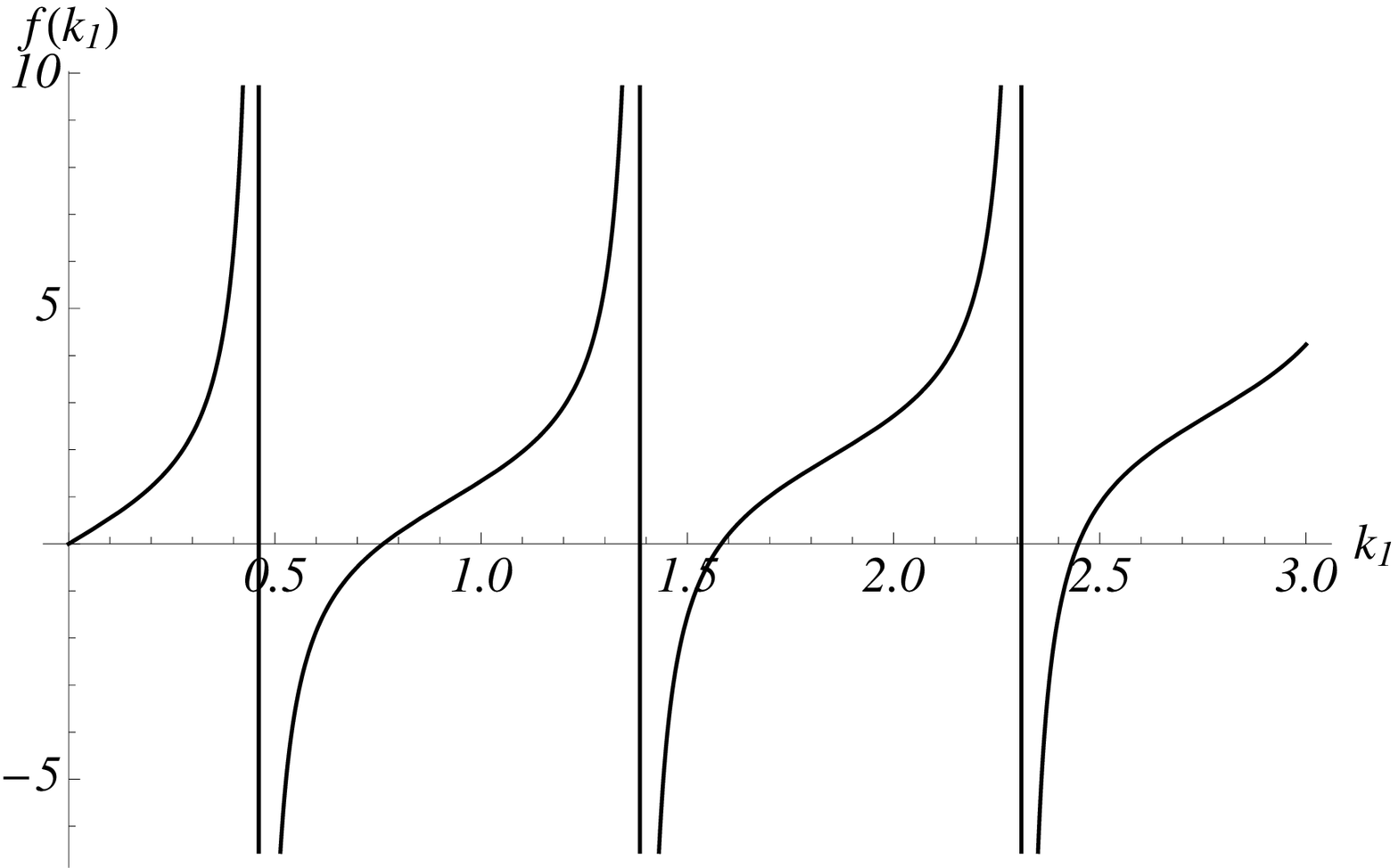}

}\subfloat[]{\includegraphics[width=0.4\columnwidth]{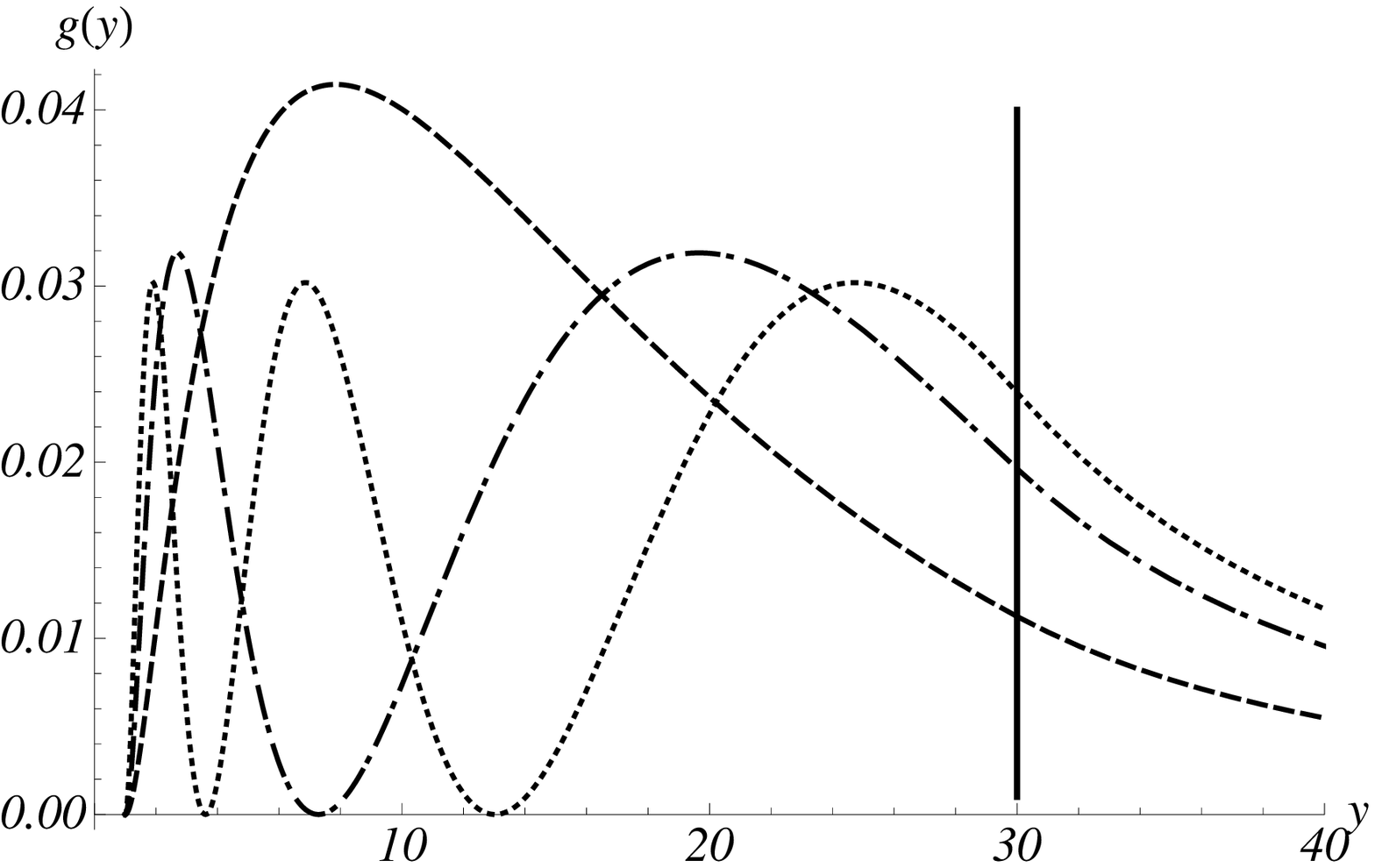}

}

\subfloat[]{\includegraphics[width=0.4\columnwidth]{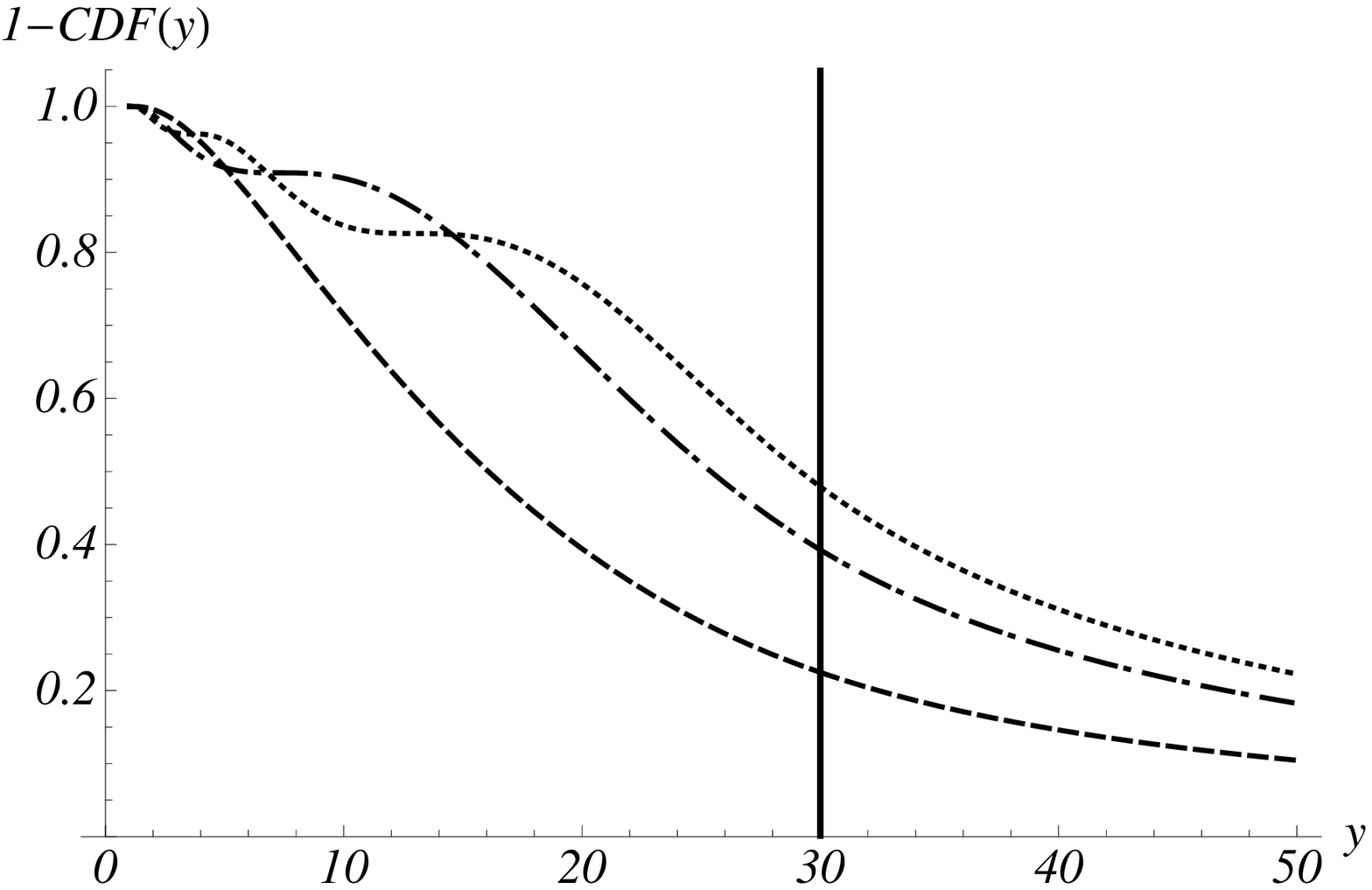}

}\subfloat[]{\includegraphics[width=0.4\columnwidth]{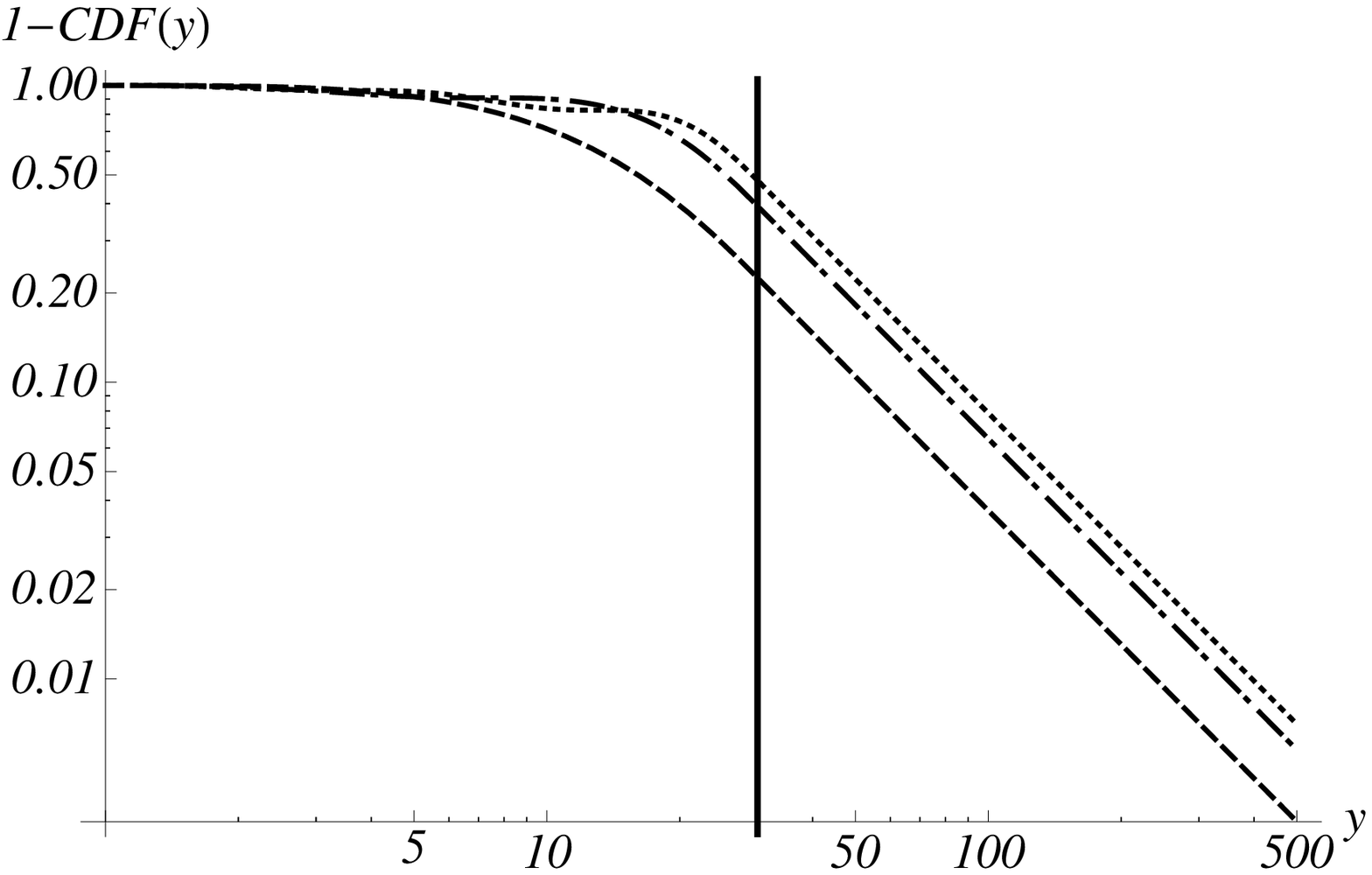}

}

\caption{(a) Plots of the shape of Eq. (\ref{eq:QuantizationFunction}); $k_{1}$
values are the zeros of this function. (b) First $3$ eigenfunctions
of $g(y)$: the ground state, dashed line; the second, dot-dashed
line and the third, dotted line. (c) The complementary CDF in a linear
plane and (d) in a Log-Log plane. The vertical line represents the
$y_{0}$ value.\label{fig:Example-Plots} }

\end{figure}

Figure (\ref{fig:Example-Plots}) shows Eq. (\ref{eq:QuantizationFunction}),
the PDF $g(y)=q(y)^{2}$ and the complementary cumulative distribution
function (CDF). Function $g(y)$ is plotted in its $3$ first eigenstates.%

\section{Dissipative term\label{sec:DissipativeTerm}}

Real systems, particularly social systems, are often conceived as
open which means they exchange matter, energy or information with
their surroundings. In this sense, a dissipative term $H(q'(y),y)$
is proposed:\begin{equation}
H(q'(y),y)=\beta\frac{q'(y)}{y}.\end{equation}
Coefficient $\beta$ will be the strength of the dissipation. This
term is constructed by analogy with the Schrödinger equation in quantum
mechanics, as shown in appendix \ref{sec:BDissipativeTerm}.

Using this dissipative term, the equation of motion becomes\begin{equation}
4q''(y)+\frac{\kappa}{t(y)^{2}}\frac{q(y)}{y^{2}}=\beta\frac{q'(y)}{y}.\end{equation}

The solutions $q_{1}(y)$ and $q_{2}(y)$ now have the general form\begin{equation}
q_{i}(y)=c_{i,1}y^{\lambda+k_{i}}+c_{i,2}y^{\lambda-k_{i}},\label{eq:SolucionSimpleDosTarifas-1}\end{equation}

where $c_{i,1}$ and $c_{i,2}$ are integration constants, and coefficient
$k_{i}$ obeys

\begin{equation}
k_{i}=\sqrt{\lambda^{2}-\frac{\kappa}{4t_{i}^{2}}},\label{eq:SimplificacionK-1}\end{equation}

with \begin{equation}
\lambda=\frac{4-\beta}{8}.\label{eq:dissipationStrength}\end{equation}

Using the same conditions as in section (\ref{sub:AParticularSolution}),
a particular solution can be written as

\begin{equation}
q(y)=\left\{ \begin{array}{lcl}
c_{1\,}y^{\lambda}\sin(k_{1}\log(y)), &  & 1\leq y\leq y_{0}\\
c_{2}\, y^{\lambda-k_{2}}, &  & y>y_{0}\end{array}\right..\label{eq:SolutionWithDissipation}\end{equation}

The scaling parameter $\alpha_{tail}$ of the Pareto distribution
can now be written in terms of the system characteristics and a dissipative
term \begin{equation}
\alpha_{tail}=2\left(k_{2}-\lambda\right)-1.\label{eq:tail-k2-lambda}\end{equation}

Using Eqs. (\ref{eq:dissipationStrength}) and (\ref{eq:tail-k2-lambda})
it can be inferred the information exchange direction with the surroundings:
if $\lambda<1/2$ the system gain information, if $\lambda>1/2$ the
system loose information and if $\lambda=1/2$ the system is in equilibrium
with the environment. In natural systems a direct equivalence with
entropy changes can be made using Fisher information as an entropy
measure \cite{Frieden1995}.

\section{application of the model}

This model can be used in a bottom-up approach, because the macroscopic
parameters of an observed distribution, such as the scaling parameter
and the $y_{min}$ value, can be related or explained using the microscopic
parameters $t_{1}$, $t_{2}$, $\kappa$ and $y_{0}$ using Eq. (\ref{eq:SimplificacionK}).
This model can also be seen in a top-down approach, because it can
be used to fit an observed PDF by means of numerical estimation of
parameters.

To fit an observed PDF, the scaling parameter $\alpha_{tail}$ and
the $y_{min}$ value can be estimated using the maximum likelihood
(ML) method as standard \cite{Clauset2009}. Then, using the corresponding
relation equation derived from the normalization condition e.g. Eqs.
(\ref{eq:k2-tail}) and (\ref{eq:Normalization}), several eigenvalues
$k_{1}$ can be computed to describe the head. Because there could
be several different values for $k_{1}$, the head can be fitted using
a linear combination of eigenvalues.

In both approaches the model allows the interpretation of the parameters
in terms of physical quantities or properties of the system. In this
sense the model is not only descriptive but might become useful to
control the system. By employing equations (\ref{eq:SimplificacionK-1})
and (\ref{eq:tail-k2-lambda}) the scaling parameter of the tail can
be adjusted, and the shape of the head can be restricted to some particular
values.

Finally, it is apparent that this model was derived using a piecewise
function $t(y)$ Eq. (\ref{eq:piecewiseF}), with only two pieces;
but the solution can be generalized to $n$ pieces. In this case the
solution will be a piecewise solution with $n$ different $q_{i}(y)$.
A PDF with several scale parameters in the tail can be useful if the
observed PDF shows more than one slope in the upper-tail.

\subsection{Fitting the real data set}

The model was used to fit $4$ data sets with power law tails. The
power law tails of selected data set were studied in \cite{Clauset2009,Newman2005}
(downloaded from http://tuvalu.santafe.edu/\textasciitilde{}aaronc/powerlaws/data.htm)
and in \cite{Lammoglia2008}, where world wealth distribution is simulated
using an agent based model (ABM).

To fit the whole PDF, first the scaling parameter $\alpha_{tail}$
and the $y_{min}$ were estimated using an ML method, then the strength
of the dissipative term and the shape of the head (a linear combination
of the first two eigenstates) were estimated by direct minimization
of the Kolmogorov-Smirnov statistic.

\begin{figure*}
\subfloat[The numbers of customers affected by electrical blackouts in the United
States between 1984 and 2002.]{\includegraphics[width=0.45\textwidth]{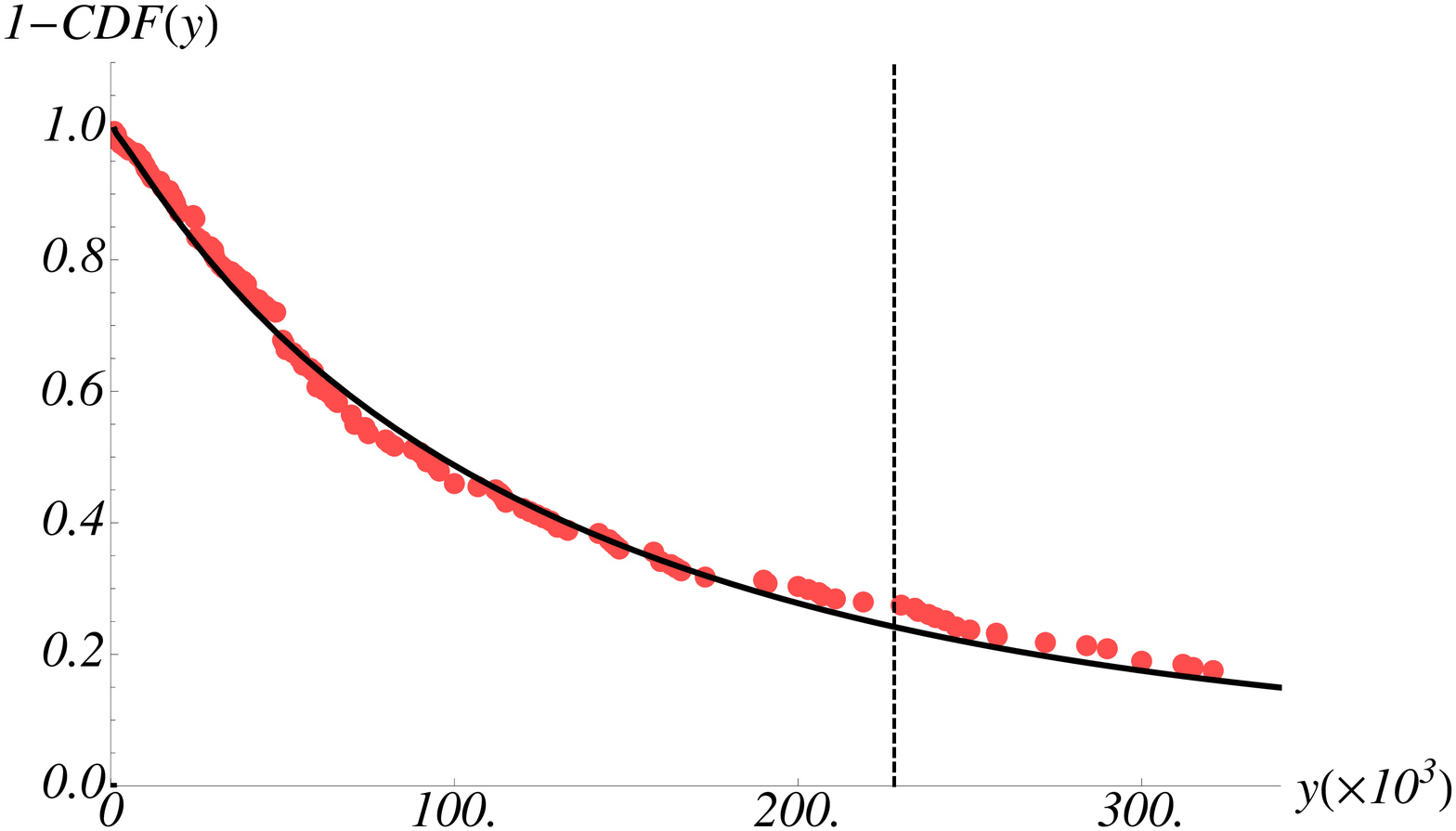}\includegraphics[width=0.45\textwidth]{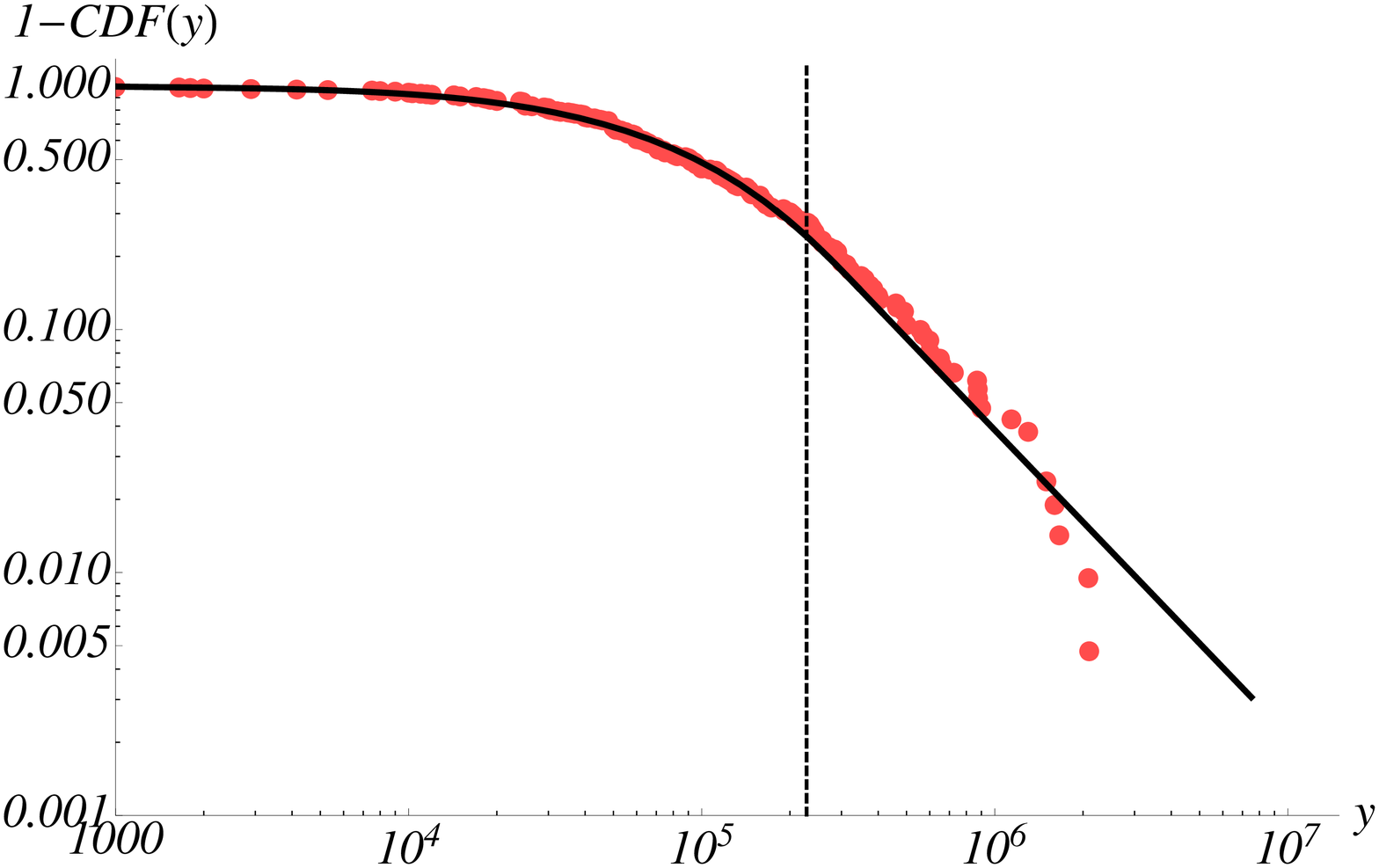}

}

\subfloat[The human population of US cities in the 2000 US Census.]{\includegraphics[width=0.45\textwidth]{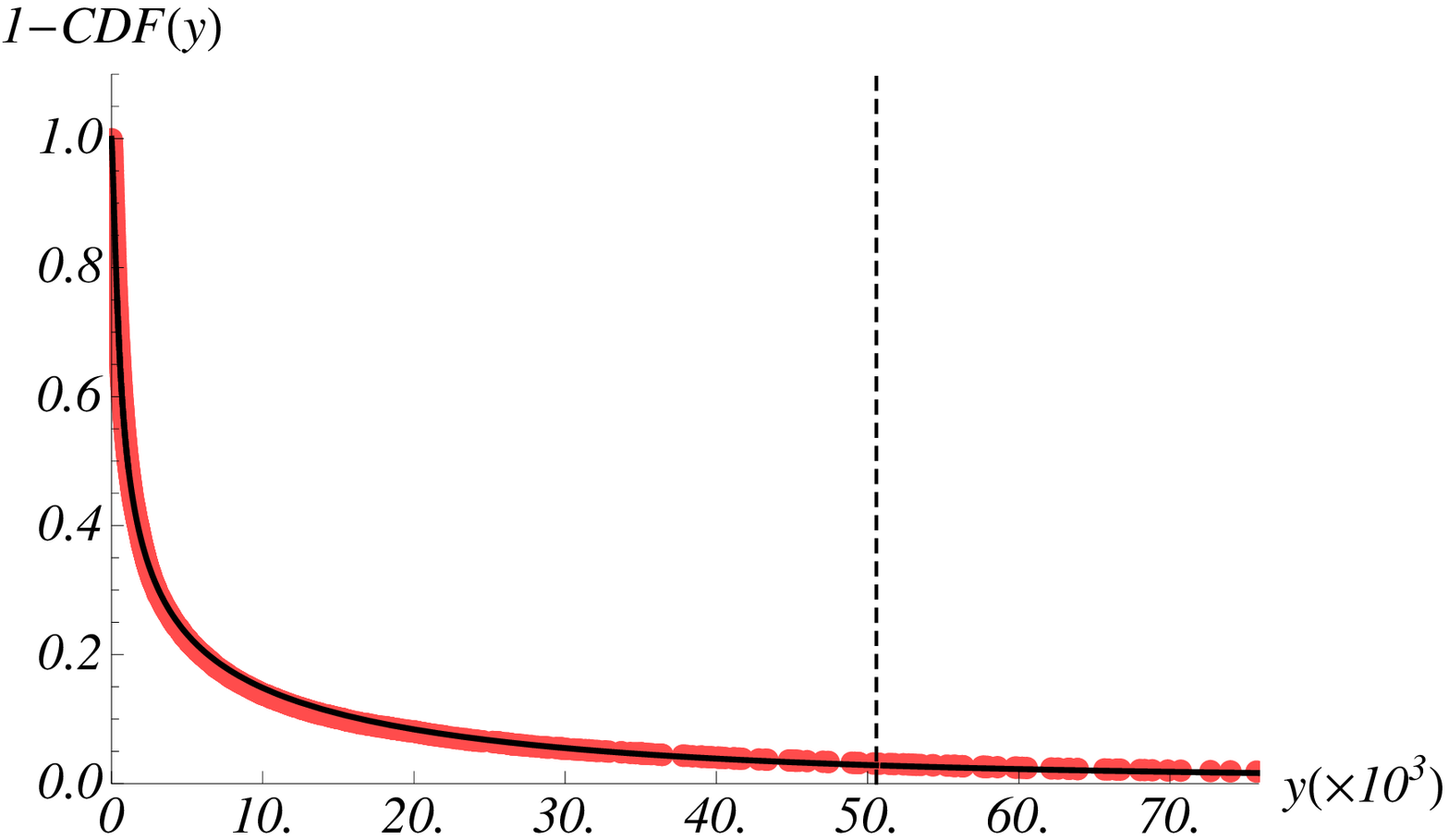}\includegraphics[width=0.45\textwidth]{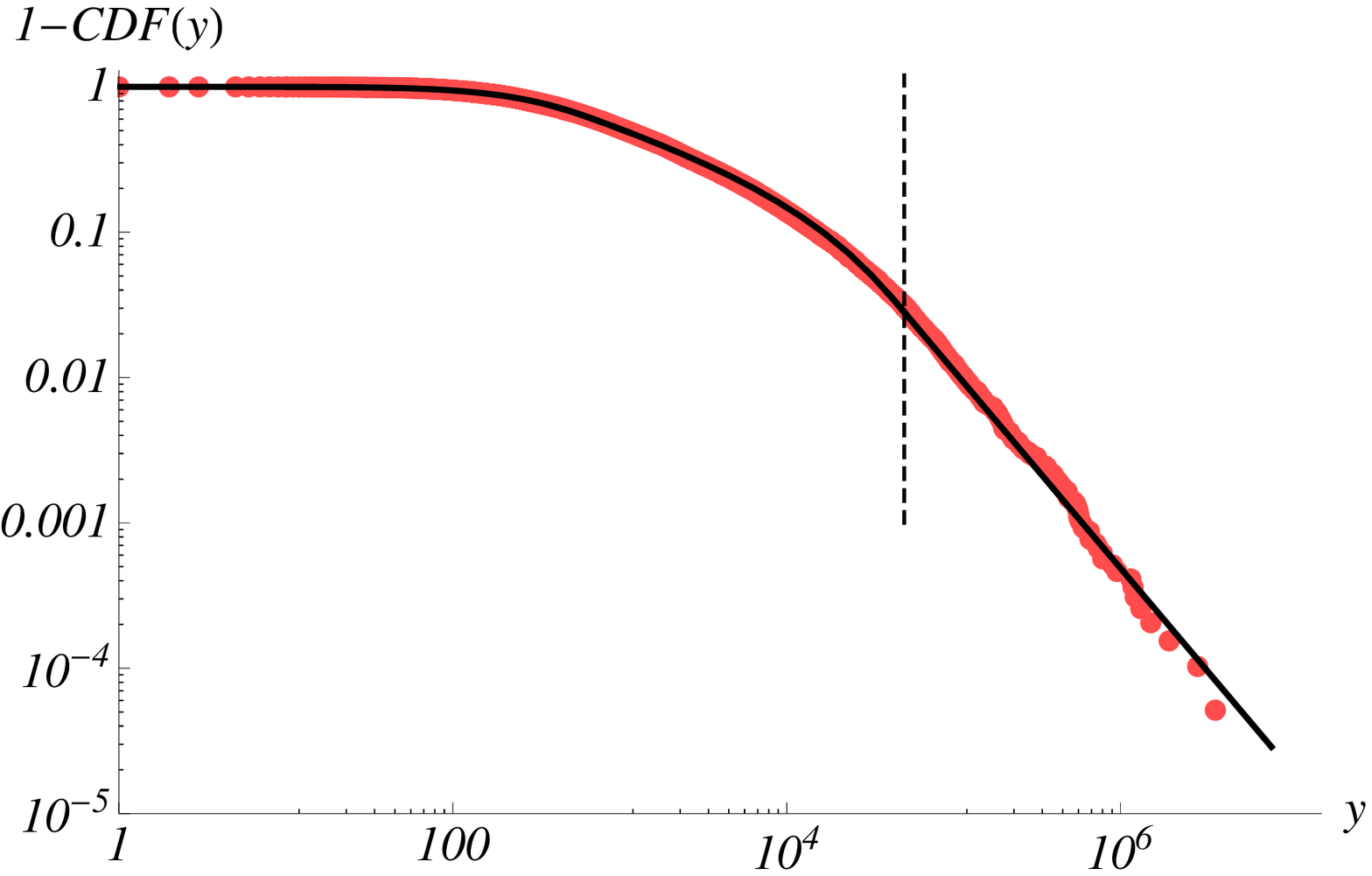}

}

\subfloat[Peak gamma-ray intensity of solar flares between 1980 and 1989.]{\includegraphics[width=0.45\textwidth]{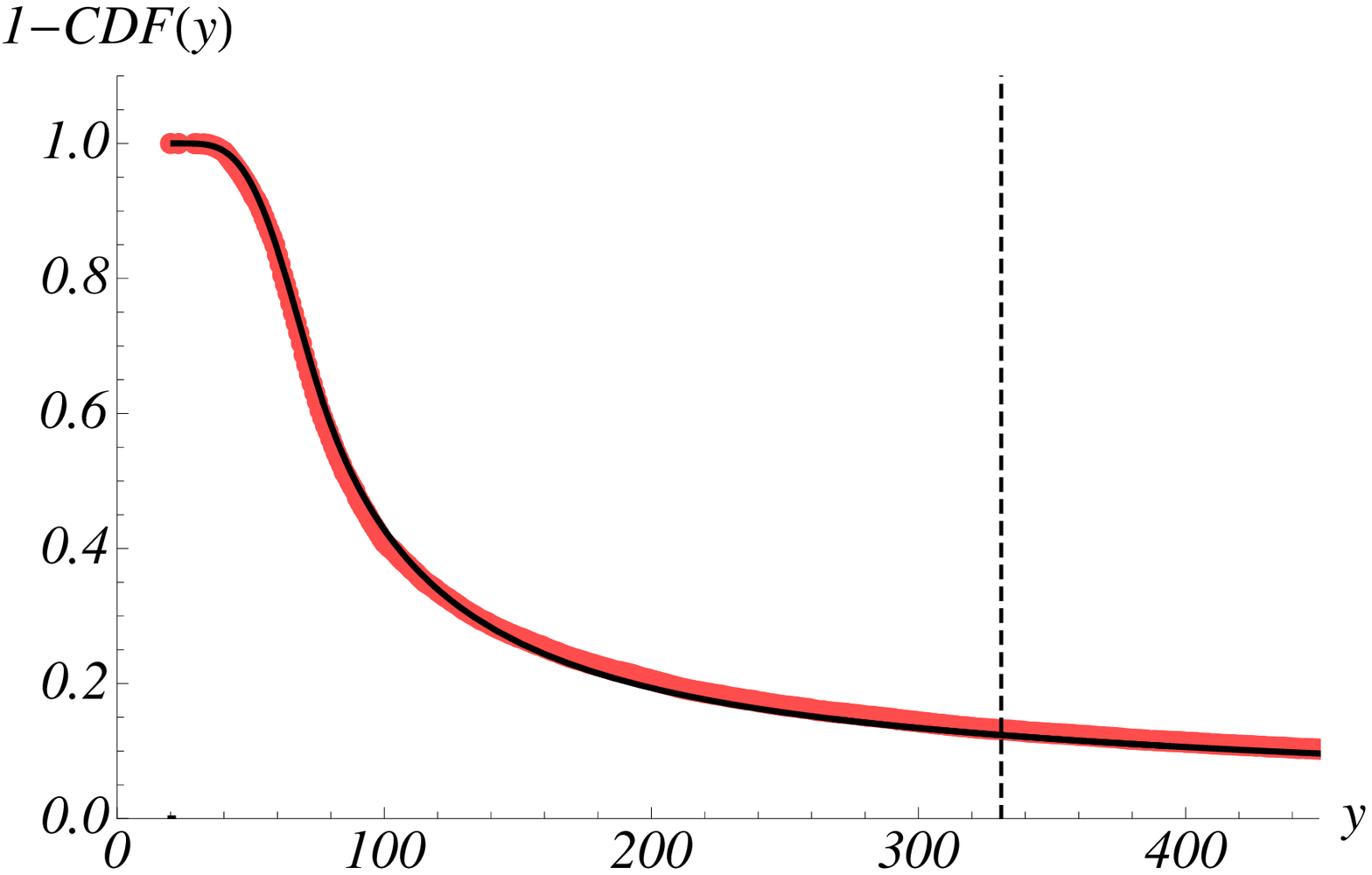}\includegraphics[width=0.45\textwidth]{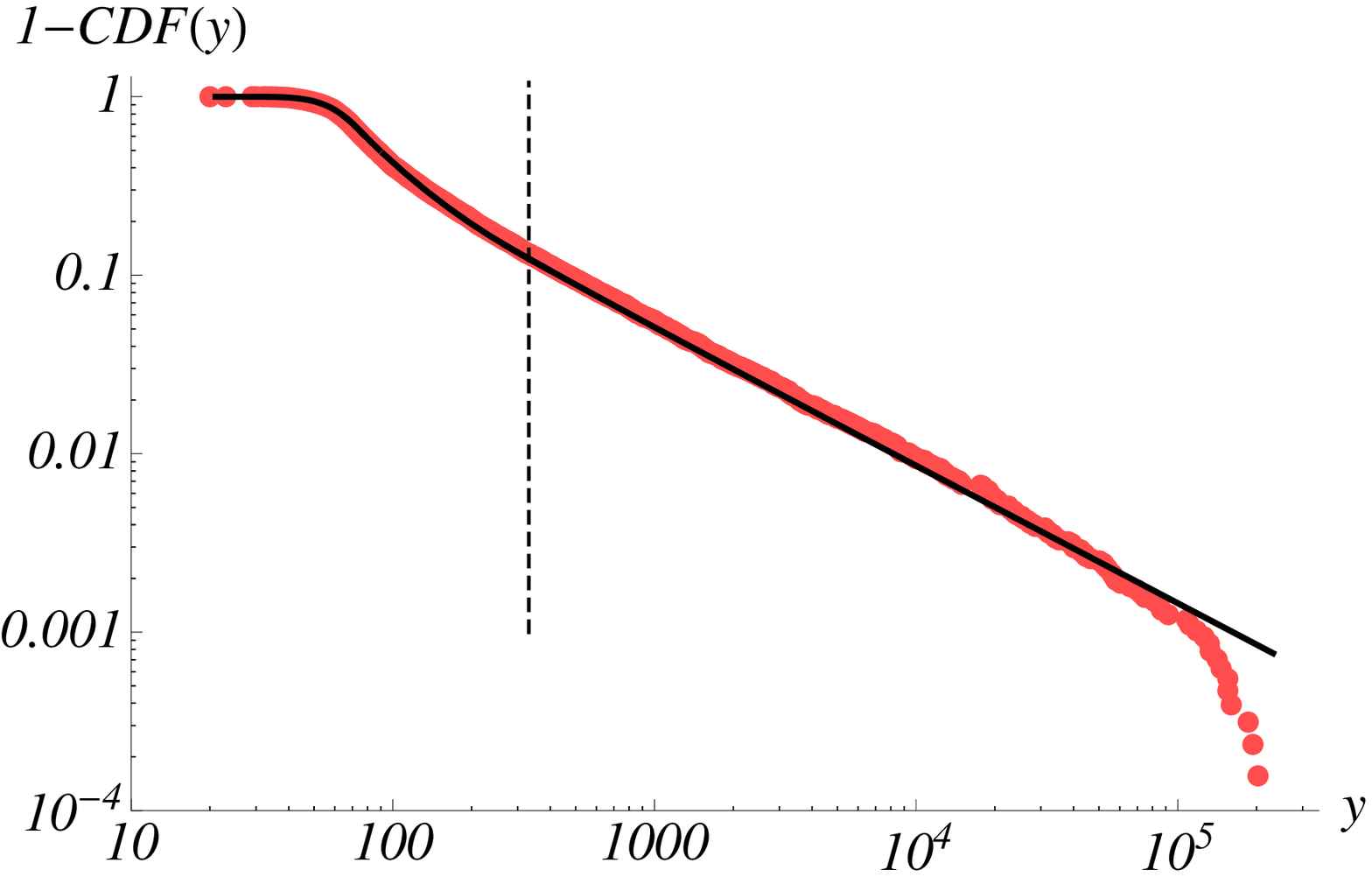}

}

\subfloat[World wealth distribution simulated using an ABM.]{\includegraphics[width=0.45\textwidth]{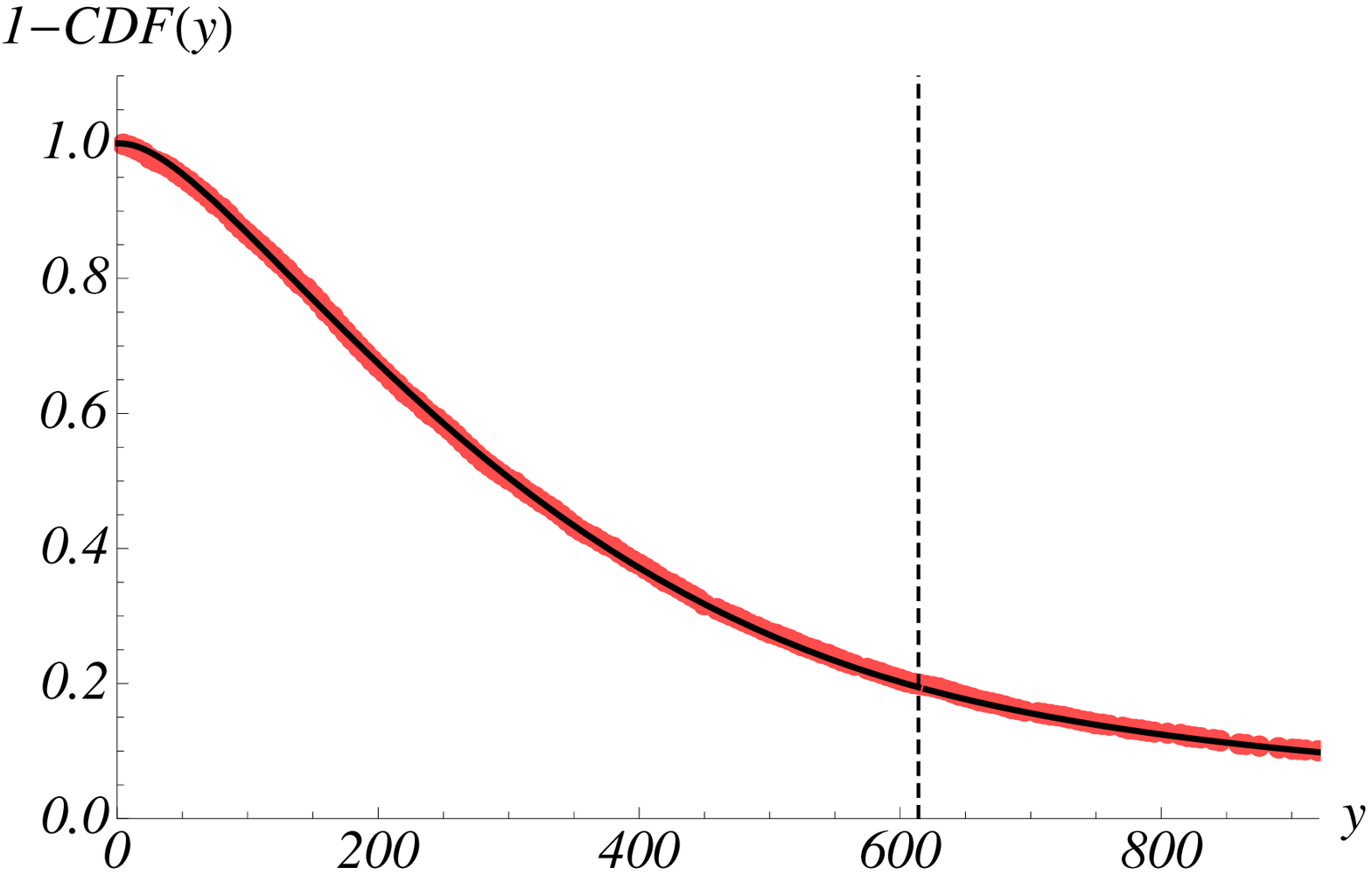}\includegraphics[width=0.45\textwidth]{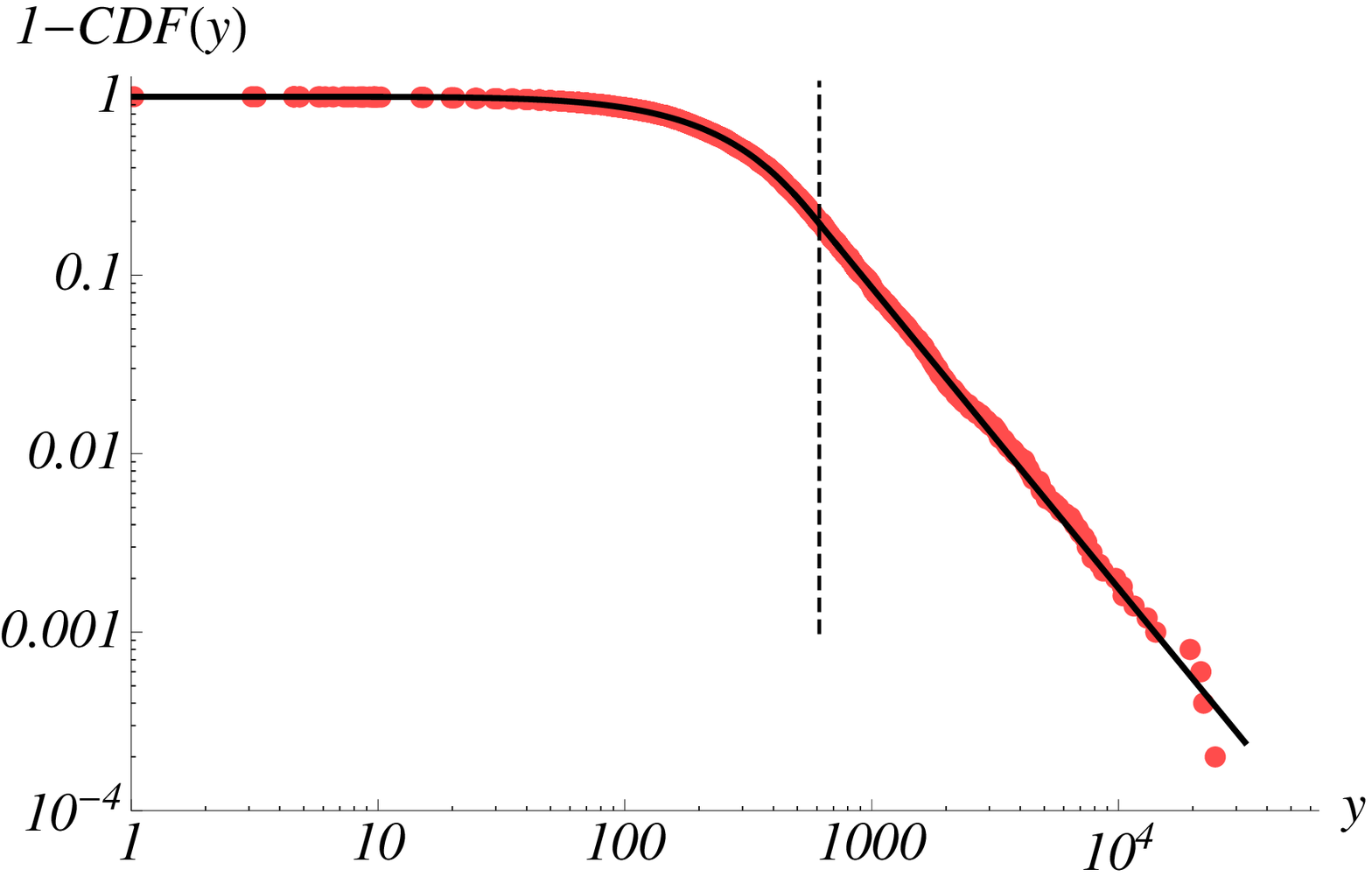}

}

\caption{Dots are the complementary CDF for each data set. The line is the
fitted model. On the left, a linear plot, on the right, a Log-Log
plot. The vertical line represents the $y_{min}$ value.\label{fig:Plotted-data-set}}

\end{figure*}

The data set comprised:
\begin{enumerate}
\item The numbers of customers affected by electrical blackouts in the United
States between 1984 and 2002 ($211$ registers).
\item The human population of US cities in the 2000 US Census ($19447$
registers).
\item Peak gamma-ray intensity of solar flares between 1980 and 1989 ($12773$
registers).
\item World wealth distribution simulated using an ABM ($5000$ agents). 
\end{enumerate}
Table (\ref{tab:Estimated-parameters}) shows boundary conditions
and estimated parameters. Figure (\ref{fig:Plotted-data-set}) shows
fitted distributions. Plots were made in a linear plane to observe
the fit of the head and in a Log-Log plane to observed the fit of
the tail.%
\begin{table*}
\begin{tabular}{|c|c|c|c|c|c|c|c|c|}
\hline 
Data set & $t(y)$ & $y_{min}$ & $\alpha_{tail}$ & $\lambda$ & Information flow & $k_{1}$ & $D_{KL}$ & $D_{C}$\tabularnewline
\hline 
1 & One step & $228000$ & $1.2(1)$ & $-0.1$ & In & $\left(0.23,\,0.46\right)$ & $0.04$ & $0.11$\tabularnewline
\hline 
2 & Two steps & $50580$ & $1.36(5)$ & $0.2$ & In & $0.41$ & $0.0099$ & $0.011$\tabularnewline
\hline 
3 & Two steps & $321$ & $0.77(2)$ & $2.0$ & Out & $0.72$ & $0.0140$ & $0.0144$\tabularnewline
\hline 
4 & One step & $614$ & $1.68(5)$ & $0.25$ & In & $\left(0.44,\,0.89\right)$ & $0.007$ & $0.023$\tabularnewline
\hline
\end{tabular}

\caption{Boundaries conditions and estimated parameters for each data set.
The one step function is a $2$ piecewise function as in Eq. (\ref{eq:piecewiseF});
the two steps functions are a $3$ piecewise function. The presence
of two values for the $k_{1}$ parameter means that the fit of the
head was made with the two first eigenstates. $D_{KL}$ is the Kolmogorov-Smirnov
statistic measured as the maximum difference between the observed
CDF and the analytical CDF. $D_{C}=1.627/\sqrt{n}$ is the Kolmogorov-Smirnov
\textit{goodness-of-fit} test critical statistic, from which the model
is rejected, so it is necessary that $D_{KL}<D_{C}$ (where $n$ is
the data length).\label{tab:Estimated-parameters}}

\end{table*}

\section{Discussion}

In the study of data set probability distribution functions, two basic
filters are usually applied. One of these involves discarding the
extreme and rare events. These events are considered to be outliers
because they do not seem to follow the general tendency. Indeed, using
Gaussian statistics it can be seen that extreme events make the variance
too wide and the results less significant \cite{Sakia1992,Cheng2005}.
As a result, the tail is excluded from the analysis. Another frequently
used filter discards the events of the head, because the main issue
is the power law tail study. The study of power law tails is an interesting
research area because properties such as self-organized criticality
and scale-free can be ascribed to the system \cite{Barabasi1999,Sornette2006,Mitzenmacher2003}.
In this kind of study, frequent events are excluded by means of a
threshold value that renders the power law tail plausible. 

In this paper a model is presented that describes the whole distribution,
involving the head and the distribution tail, and no data is discarded.
Moreover, a relation between the scaling parameter of the power law
tail and the shape of the general tendency of the head is also derived. 

The model constructed becomes interesting for the study of many systems
that exhibit power law tails, because it proposes an equation of motion
that describes the behavior of the system and allows its parameters
to be interpreted. Even if the model has a piecewise restriction function
Eq. (\ref{eq:piecewiseF}) it can be used to explain and fit several
systems, in particular social systems. A lot of social variables,
such as prices, taxes, incentives, budgets, etc., follow piecewise
restrictions. The model can also be used to gain insight concerning
microscopic processes in natural systems that follow power law tails
and are not already well understood

The analysis and fitting of power law tails is an active research
area. The computation of the scaling parameter of the PDF and the
$y_{min}$ value (value from where the power law is plausible) is
not as trivial as it seems \cite{Clauset2009}. The model developed
here contributes to this research area: first, by taking into account
all the registers in the data set; second, by offering a new interpretation
of the scaling parameter and third, by providing a relation between
the tail and the head.

A dissipative term was introduced into the model. This term permits
the description of a large number of phenomena, and it should be interpreted
in terms of the particular scenario of the system under study. For
now, a simple interpretation is given in terms of the flow direction
of the dissipation. A gain of information can be interpreted, as order
is been established on the system, i.e. data sets $1$, $2$ and $4$;
a loss of information can be interpreted as the system seeking its
thermodynamic equilibrium state, i.e. data set $3$.

\section{Conclusions}

Using EPI theory according to the basic information principle, the
equation of motion and the system PDF with a power law tail are deduced.
The calculated PDF is as useful in describing the tail as the head
of the PDF. In particular, the head is described as a combination
of several eigenstates. This model, in a bottom-up approach, gives
an explanation of the macroscopic parameters, $\alpha_{tail}$ and
$y_{min}$, and the shape of the head, from the microscopic parameters
of the system. In a top-down approach the model is equally useful
in fitting the tail as the head of an observed PDF, giving an interpretation
for the estimated parameters.

The principal contribution of this paper to the analysis of systems
with power law tail behavior concerns the relation between rare events
in the tail and frequent events in head. In the literature, no one
denies that there must be a relation between the head and the tail;
here, a relation is proposed, by means of eigenstates. With this model
it is not necessary to filter the data set and discard events: it
is possible to study the general behavior in the head and take into
account that the system presents complex characteristics as a power
law in the tail.

Not only will the proposed dissipative term give a better fit in a
top-down approach, but it permits a description to be made of the
interactions between the systems and their surroundings. Using this
model the scaling parameter in the tail could be understood as a combination
of the structure and properties of a system and its interaction with
the environment. 

We believe that our model is capable of taking into account the fact
that in many systems behavior is a mixture of order and disorder.
It is also possible to gain insights regarding the plausible causes
of emergent behavior, given the way parameters are fed into the model. 

\appendix

\section{Bounded Information\label{sec:ABoundedInfo}}

To write the information functional

\begin{equation}
\mathcal{F}=\int dy\,\frac{g'(y)^{2}}{g(y)}-\kappa\int dz\,\frac{h'(z)^{2}}{h(z)},\label{eq:Functional-FI-1}\end{equation}

in terms of $g'(y)$, $g(y)$ and $y$, the term

\begin{equation}
I^{(z)}=\int dz\,\frac{h'(z)^{2}}{h(z)},\end{equation}
must be re-written.

The conservation of the probability is applied, using Eq. (\ref{eq:relacion-lineal})
and (\ref{eq:piecewiseF}). Supposing that $y$ and $t(y)$ are random
variables the PDF $h(z)$ can be written as the product of two random
variables; by definition

\begin{eqnarray}
h(z=t(y)\, y) & = & \int dy\,\frac{1}{|y|}g(y)p(\frac{z}{y}|y),\label{eq:MultiplicaionVariables}\end{eqnarray}
where the function $p(t|y)$ is the probability of $t=t_{1}$ or $t=t_{2}$,
given $y$. This probability function, $p(t|y)$, can be written in
term of the delta function $\delta(y-y_{0})$ and the Heaviside function
$H(y-y_{0})$ as

\begin{eqnarray}
p(t|y) & = & a_{1}\delta(t-t_{1})(1-H(y-y_{0})\nonumber \\
 &  & +a_{2}\delta(t-t_{2})H(y-y_{0}),\label{eq:Prob-Tarifa}\end{eqnarray}

where $a_{1}+a_{2}=1$.

Using equations Eq. (\ref{eq:Prob-Tarifa}) in Eq. (\ref{eq:MultiplicaionVariables})
we get

\begin{widetext}

\begin{eqnarray}
h(z) & = & \intop_{0}^{\infty}dy\,\frac{1}{|y|}g(y)\left\{ a_{1}\delta(\frac{z}{y}-t_{1})(1-H(y-y_{0})+a_{2}\delta(\frac{z}{y}-t_{2})H(y-y_{0})\right\} \nonumber \\
 & = & \intop_{0}^{y_{0}}dy\,\frac{1}{|y|}g(y)a_{1}\delta(\frac{z}{y}-t_{1})+\intop_{y_{0}}^{\infty}dy\,\frac{1}{|y|}g(y)a_{2}\delta(\frac{z}{y}-t_{2}).\label{eq:Pc}\end{eqnarray}

Now $h'(y)$ is calculated taken the derivative $\frac{\partial}{\partial y}$
of Eq. (\ref{eq:Pc}),

\begin{eqnarray}
\frac{\partial h(z)}{\partial z} & = & \intop_{0}^{y0}dy\,\frac{1}{|y|}g(y)a_{1}\,\frac{\partial}{\partial z}\delta(\frac{z}{y}-t_{1})+\intop_{y_{0}}^{\infty}dy\,\frac{1}{|y|}g(y)a_{2}\,\frac{\partial}{\partial z}\delta(\frac{z}{y}-t_{2})\nonumber \\
 & = & -\intop_{0}^{y0}dy\,\frac{1}{|y|}g(y)a_{1}\,\frac{\left|y\right|}{z}\delta(z-t_{1}y)-\intop_{y_{0}}^{\infty}dy\,\frac{1}{|y|}g(y)a_{2}\,\frac{\left|y\right|}{z}\delta(z-t_{2}y)\nonumber \\
 & = & -\frac{1}{z}h(z).\label{eq:DeduccionPpc}\end{eqnarray}

\end{widetext}

This last development takes into account that $g(\frac{z}{t_{i}})=t_{i}h(z)$.

Finally, using Eqs. (\ref{eq:DeduccionPpc}) and (\ref{eq:relacion-lineal}),
we write $I^{(z)}=\int dz\,\frac{h'(z)^{2}}{h(z)}$ as

\begin{eqnarray}
I^{(z)} & = & \int dz\,\frac{h(z)}{z^{2}}\nonumber \\
 & = & \intop_{1}^{z^{*}}dz\,\frac{h(z)}{z^{2}}+\intop_{z**}^{\infty}dz\,\frac{h(z)}{z^{2}}\nonumber \\
 & = & \intop_{1}^{y_{0}}dy\, t_{1}\frac{h(t_{1}y)}{t_{1}^{2}y^{2}}+\intop_{y_{0}}^{\infty}dy\, t_{2}\frac{h(t_{2}y)}{t_{2}^{2}y^{2}}\nonumber \\
 & = & \int dy\,\frac{1}{t(y)^{2}}\frac{g(y)}{y^{2}},\label{eq:FI-c-1}\end{eqnarray}

where $z*=t_{1}y_{0}$ and $z**=t_{2}y_{0}$, and $h(t_{i}y)=\frac{1}{t_{i}}g(y)$.

\section{Dissipative term\label{sec:BDissipativeTerm}}

To introduce a dissipative term into the analysis it is only necessary
to equate the second order differential equation Eq. (\ref{eq:EqMotion})
to a function $H(q'(y),y)$, where $H(q'(y),y)$ has a general form
of a dissipative term, that is, in terms of $q'(y)$. Here, a particular
form for the dissipative term is inspired by the mathematical equivalence
of Eq (\ref{eq:EqMotion}) with the Schrödinger equation. Under the
change of variable $\tilde{y}=\log\, y$, the Schrödinger equation
of a free particle

\begin{equation}
a\varphi''(\tilde{y})+b\varphi(\tilde{y})=0,\label{eq:EQ-tipo-Schr}\end{equation}

becomes

\begin{equation}
aq''(y)+a\frac{q'(y)}{y}+b\frac{q(y)}{y^{2}}=0.\label{eq:EQ-EPI}\end{equation}

In the same way, the general solution of the Schrödinger equation
\begin{equation}
\varphi(\tilde{y})=A\exp\left(k\tilde{y}\right)+B\exp\left(-k\tilde{y}\right),\label{eq:SOL-tipo-Schr}\end{equation}

becomes

\begin{equation}
q(y)=A\, y^{k}+B\, y^{-k}.\label{eq:SOL-EPI}\end{equation}

In both cases, $k=\pm\sqrt{-b/a}$. 

Comparing Eq. (\ref{eq:EQ-EPI}) with Eq. (\ref{eq:EqMotion}) an
additional term $a\frac{q'(y)}{y}$ can be identified. Because this
term has a first derivative, by analogy with dissipative forces we
propose that $a\frac{q'(y)}{y}$ can be related to a dissipative source
of information. So, by analogy with a quantum system we propose

\begin{equation}
H(q'(y),y)=\beta\frac{q'(y)}{y},\label{eq:Disipacion}\end{equation}
where, $\beta$ is a constant, related to the strength of the dissipation.

The dissipative term is still not well understood, but allows a sharp
estimate to be made of the scaling parameter at the tail: see Eq.
(\ref{eq:SolutionWithDissipation}). It is also interesting to discuss
a new interpretation for the scaling parameter of the power law tail
using Eq. (\ref{eq:tail-k2-lambda}). 
\begin{acknowledgments}
We acknowledge COLCIENCIAS\textquoteright{}s financial support R.B,
and the Research Fund of the Engineering Faculty of the Universidad
de los Andes, Colombia R.Z..
\end{acknowledgments}
\bibliographystyle{apsrev4-1}
\bibliography{PaperThModel}

\end{document}